\documentclass[aps,prb,reprint]{revtex4-2}

\usepackage{graphicx}
\usepackage{bm}
\usepackage{bbold}
\usepackage{amssymb}
\usepackage{amsmath}
\usepackage{natbib}

\usepackage{color}

\renewcommand{\vec}[1]{{\mathbf #1}}

\bibpunct{[}{]}{,}{n}{}{}

\begin{document}

\title{Anderson localization of light by impurities in a solid transparent matrix}

\author{S.E. Skipetrov}
\email[]{Sergey.Skipetrov@lpmmc.cnrs.fr}
\affiliation{Univ. Grenoble Alpes, CNRS, LPMMC, 38000 Grenoble, France}

\author{I.M. Sokolov}
\email[]{igor.m.sokolov@gmail.com}
\affiliation{Peter the Great St. Petersburg Polytechnic University, 195251 St. Petersburg, Russia}

\date{\today}

\begin{abstract}
A solid transparent medium with randomly positioned, immobile impurity atoms is a promising candidate for observation of Anderson localization of light in three dimensions. It can have low losses and allows for mitigation of the detrimental effect of longitudinal optical fields by an external magnetic field, but it has its own issues: thermal oscillations of atoms around their equilibrium positions and inhomogeneous broadening of atomic spectral lines due to random local electric fields. Our calculations suggest  that these complications should not impede observation of Anderson localization of light in such materials provided that sufficiently high number densities of impurities can be reached. The thermal oscillations hardly affect light propagation whereas the inhomogeneous broadening can be compensated for by increasing the number density of impurities.
\end{abstract}

\maketitle

\section{Introduction}
Anderson localization---a complete halt of wave transport due to disorder \cite{anderson58}---turns out to be difficult \cite{wiersma97,scheffold99,wiersma99,beek12,storzer06,
sperling13,scheffold13,maret13,beek12,sperling16,skip16njp} and likely even impossible \cite{yamilov23} to reach for light in fully disordered three-dimensional (3D) dielectric media. Partially ordered structures such as disordered photonic crystals \cite{john87,john91,john93} or hyperuniform materials \cite{haberko20,scheffold22}
with a photonic band gap
in their optical spectrum may
feature
spatially localized optical modes
but only
at frequencies near a band edge where the optical density of states is strongly suppressed. Metallic structures may be better suitable for observation of Anderson localization of light than dielectric ones \cite{genack91,yamilov23,yamilov24} although real metals suffer from significant losses that can make optical experiments difficult to conduct and interpret  \cite{genack91,chabanov00}.

Cold atoms exhibiting purely elastic, lossless scattering have been proposed as an alternative medium for observation of Anderson localization of light in 3D
\cite{kaiser00,kaiser09}
but later longitudinal electric fields have been predicted to prevent such an observation \cite{skip14prl,tiggelen21}. A possible solution to this problem consists in placing the atoms in a strong external magnetic field that partially suppresses longitudinal fields \cite{skip15prl,skip18prl}. Theoretical work has predicted the expected signatures of Anderson localization for coherent laser light in optically thick and dense cold-atom ensembles placed in a strong external magnetic field: slowing down of the temporal decay of a transmitted pulse \cite{skip16pra}, enhanced fluctuations of scattered intensity \cite{cottier19}, step-like profile of average intensity inside the atomic sample \cite{skip19prl}. However, the experimental realization of these theoretical predictions is currently impeded by the following two main obstacles. First, it is difficult to prepare cold-atom samples that would be optically thick to ensure multiple scattering (size $L >$ photon mean free path $\ell$) and, at the same time, sufficiently dense to reach Anderson localization (atomic number density $\rho > \lambda_0^{-3}$, where $\lambda_0$ is the resonance wavelength in the free space) \cite{kaiser05,kaiser09}.  Second, experiments are necessarily performed at a low but
nonzero
temperature (typically, $T \sim 100$ $\mu$K \cite{kaiser05,guerin16}) leading to residual motion of atoms that
washes out interference phenomena for quasi-resonant light already in the weak localization regime \cite{kupriyanov04,labeyrie06} and is expected to be particularly detrimental for Anderson localization \cite{skip16pra,kuraptsev20}.
This is qualitatively similar to the breakdown of localization of electrons in disordered solids due to interaction with thermal phonons \cite{mott68,shklovskii84}.

A transparent solid medium with impurity atoms or ions embedded at random locations is an alternative physical system described by the theoretical model of immobile point-like scattering centers (atoms or ions) developed in Refs.\ \cite{skip14prl,tiggelen21,skip15prl,skip16pra,skip18prl}.
Well-known examples of such materials are uranium glass (with U atoms in, e.g., oxide diuranate form as impurities) \cite{hahlweg12}---which is, by the way, the first optical material in which optical nonlinearity has been observed \cite{wawilow26}---and ruby (with Cr$^{3+}$ ions as impurities), for which the possibility of Anderson localization
has been already evoked \cite{koo75,chu80,jessop82}.
Another example is the diamond crystal with multiple nitrogen-vacancy (NV) centers---a system with promising applications in quantum information science \cite{aharonovich11,doherty13}.
Scattering properties of individual, isolated NV centers have been studied \cite{tran17} and they have been used as probes of strong scattering in random media \cite{rui10}. Transparent solids with embedded impurity atoms have a number of advantages as compared to cold atomic gases as far as the observation of Anderson localization of light is concerned. On the one hand, they can be found in nature or fabricated without fundamental limitations on the
sample
size $L$ or impurity number density $\rho$, so that the conditions $L > \ell$ and $\rho > \lambda_0^{-3}$ can be fulfilled. On the other hand, experiments can be performed at room or, in any case, not-too-low temperature because impurities are fixed at their positions in the matrix and do not fly away. Two problems arise nevertheless in these systems. First, local electric fields induce position-dependent shifts of resonance frequencies of impurity atoms, driving them out of resonance with each other. Such an inhomogeneous broadening of the spectrum is expected to reduce the maximum achievable  scattering strength, which may bring the system away from the Anderson localization regime. Second, the problem of atomic motion that is crucial in atomic gases, does not disappear completely here either: the solid matrix still allows for oscillations of impurity atoms about their equilibrium positions with an amplitude and a frequency determined by the temperature. At the first glance, such oscillations could be expected to be less detrimental for Anderson localization of light than the free atomic motion, but it is not clear to which degree. Thus, it is not clear \textit{a priori} whether the replacement of free atomic motion by oscillations brought about by embedding the atoms in a solid host medium overweighs  the detrimental impact of random local fields due to the very same host medium and makes Anderson localization achievable under realistic conditions.

In the present paper we analyze a theoretical model of light scattering including both oscillations of impurity atoms about their equilibrium positions and random frequency shifts of atomic energy levels. Our analysis suggests that Anderson localization of light should be achievable in random ensembles of impurity atoms embedded in a transparent solid host medium under realistic experimental conditions and, in particular, at finite temperatures. Fast but small-amplitude oscillations of impurity atoms around their equilibrium positions have virtually no effect on light propagation, in analogy with Dicke narrowing of Doppler broadened spectral lines in the presence of atomic collisions  \cite{dicke53}. In contrast, the inhomogeneous broadening of atomic spectral lines does suppress scattering and makes Anderson localization more difficult to reach. However, this suppression can be compensated by increasing the number density of atoms, so that a photon mobility edge always exists, whatever the broadening.

\section{Model}
We start with a standard Hamiltonial $\hat{H}$ describing the free electromagnetic field coupled with $N$ identical two-level atoms located at positions $\{ \vec{r}_n \}$ and having each a ground state $|g\rangle$ with energy $E_g$ and angular momentum $J_g = 0$, three degenerate excited states $|e_m\rangle$, $m = 0,\pm1$, with energy $E_e = E_g + \hbar\omega_0$, spontaneous decay rate $\Gamma_0$, and angular momentum $J_e = 1$, and
dipole moments $\vec{d}_{e_m g}$ of the transitions $| g\rangle \to |e_m \rangle$ \cite{cohen92,morice95}.
Symmetry considerations impose $|\vec{d}_{e_mg}| = d$ independent of $m$. We follow the now well-known procedure to eliminate field variables and obtain an effective non-Hermitian Hamiltonian $\hat{H}_{\text{eff}}$ describing immobile atoms coupled by the quasi-resonant electromagnetic field \cite{lehmberg70,agarwal70}. When the lifting of the degeneracy of the excited state by an external magnetic field $\vec{B} = B \vec{e}_z$ and the inhomogeneous broadening of atomic spectral lines $\omega_n = \omega_0 + \Delta_n$ are included in the model, the effective Hamiltonian is composed of $N \times N$ blocks of size $3 \times 3$ that in units of $\hbar\Gamma_0/2$ can be written as
\begin{eqnarray}
\begin{aligned}
\left( \hat{H}_{\text{eff}} \right)_{n n'} &= \left[\left( i - \frac{2\Delta_n}{\Gamma_0} \right) \mathbb{1}
- \frac{2 \Delta_B}{\Gamma_0}
\begin{pmatrix}
-1 & 0 & 0 \\
0 & 0 & 0 \\
0 & 0 & 1
\end{pmatrix}
\right]
\delta_{n n'}
\\
&-
\frac{6\pi}{k_0} (1 - \delta_{n n'})
\hat{d} \hat{G}(\vec{r}_n-\vec{r}_{n'}) \hat{d}^{\dagger}
\end{aligned}
\label{heff}
\end{eqnarray}
where $\Delta_B = g_e \mu_B B/\hbar$ is the Zeeman shift, $\mu_B$ is the Bohr
magneton, $g_e$ is the Land\'{e} factor of the excited state, $\mathbb{1}$ is the $3 \times 3$ identity matrix,
\begin{eqnarray}
\hat{G}(\vec{r}) = -\frac{e^{i k_0 r}}{4 \pi r}
\left[ P(i k_0 r) \mathbb{1}
+ Q(ik_0 r) \frac{\vec{r} \otimes \vec{r}}{r^2} \right]
\label{green}
\end{eqnarray}
is the dyadic Green's function of Maxwell equations with $k_0 = \omega_0/c = 2\pi/\lambda_0$, $P(u) = 1 - 1/u + 1/u^2$, $Q(u) = -1 + 3/u - 3/u^2$. The matrix
\begin{eqnarray}
{\hat d} =
\begin{pmatrix}
1/\sqrt{2} & i/\sqrt{2} & 0 \\
0 & 0 & 1 \\
-1/\sqrt{2} & i/\sqrt{2} & 0
\end{pmatrix}
\label{dmatrix}
\end{eqnarray}
converts the Green's tensor from the Cartesian coordinate basis to the so-called cyclic one \cite{varshalovich88,skip15prl}.
Following previous works \cite{skip14prl,tiggelen21,skip15prl,skip16pra,skip18prl}, we restrict our consideration to states in which no more than one excitation (photon) is present in the system, which corresponds to the one-particle setting suitable for discussing the Anderson localization phenomenon.

\begin{figure}[t]
\hspace*{5mm}
\includegraphics[width=\columnwidth]{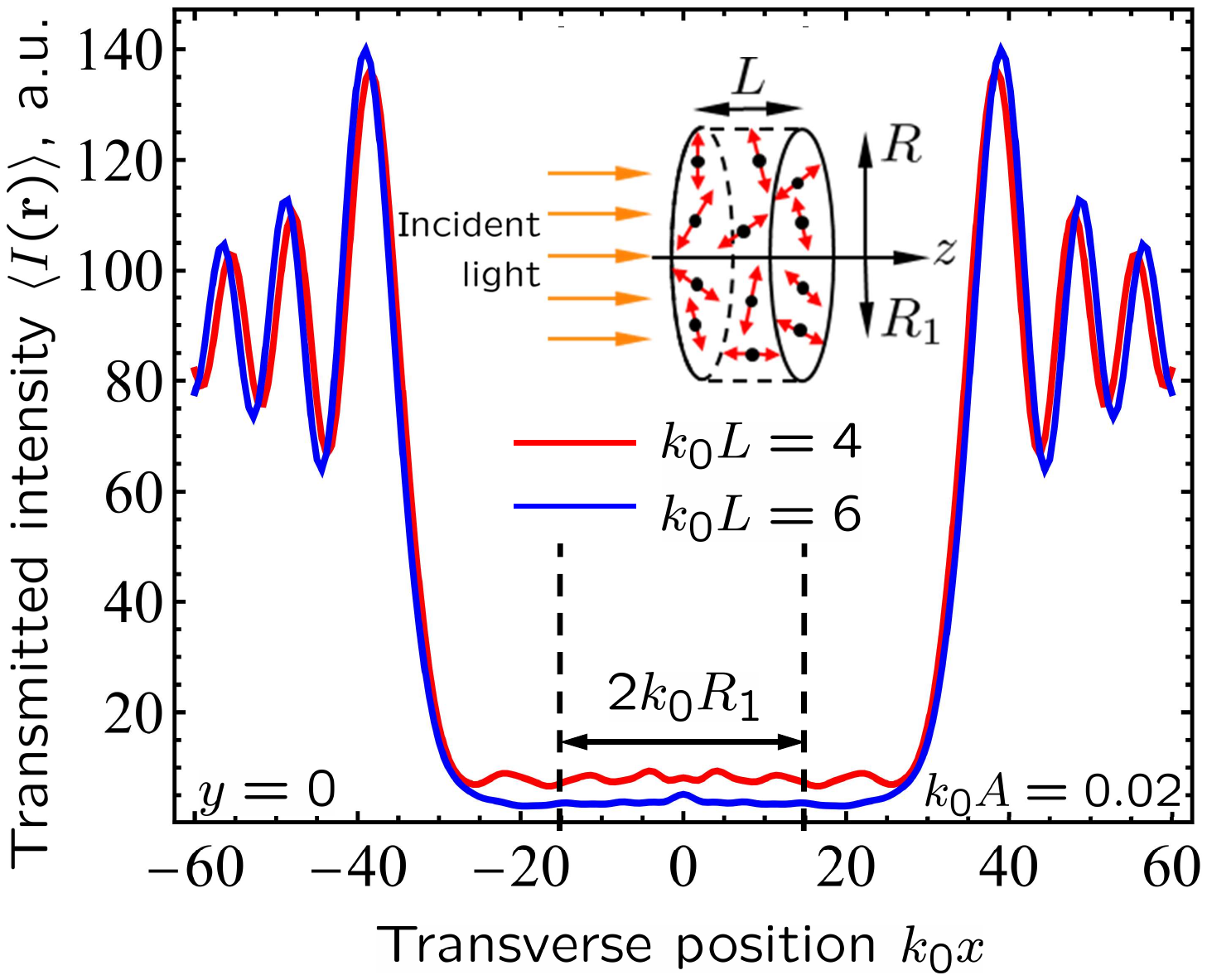}
\vspace*{-7mm}
\caption{\label{fig_trans}
Typical ensemble- and time-averaged transmission profiles obtained in the plane $z = L + 12/k_0$ for a plane wave incident on a cylindrical sample of radius $k_0 R = 30$ shown in the inset where impurity atoms are represented by black dots while their fast oscillations around random equilibrium positions are symbolized by red arrows.
$\rho/k_0^3 = 0.2$,
$\Delta_0 = 0$, $\delta \omega = 0.8\Gamma_0$, $k_0 A = 0.02$ and $\Omega = 30\Gamma_0$
for this figure. Vertical dashed lines indicate the range of spatial averaging to obtain $\langle T_a \rangle$ that would approximate the result for an infinitely wide sample ($R \to \infty$), see Eq.\ (\ref{intaver}).}
\end{figure}

To study the optical transport through the system described by the Hamiltonian (\ref{heff})
with time-dependent positions of impurity atoms, we allow for slow evolution of $\{ \vec{r}_n \}$ with time inside a
solid, transparent cylinder of radius $R$ and
thickness $L \ll R$,
see the inset of Fig.\ \ref{fig_trans}.
We model an experiment in which the sample is illuminated by a monochromatic plane wave $\vec{E}_0(\vec{r}) \exp(- i\omega t) = \vec{u}_0 E_0 \exp(i k z - i\omega t)$ incident on its face $z=0$. The vector $\bm{\beta}_n = \{\beta_{nm} \}$ of probability amplitudes for the atom $n$ to be excited in a state $|e_m\rangle$ obeys \cite{courteille10,sokolov11,kuraptsev20}
\begin{eqnarray}
\begin{aligned}
\frac{\partial \bm{\beta}_{n}(t)}{\partial t} &=
\left[ i \left( \delta\omega - \Delta_n \right) - \frac{\Gamma_0}{2} \right] \bm{\beta}_{n}(t) + \frac{i d}{2\hbar} \hat{d} \vec{E}_0(\vec{r}_n)
\\
&- i \frac{3\pi \Gamma_0}{k_0}
\sum\limits_{n' \ne n}^{N}
\left\{ \hat{d} \hat{G}\left[
\vec{r}_n(t)-\vec{r}_{n'}(t)
\right] \hat{d}^{\dagger} \right\} \bm{\beta}_{n'}(t)
\end{aligned}
\label{dynamic}
\end{eqnarray}
where $\delta\omega = \omega-\omega_m$ is the detuning of the incident light with respect to one of the Zeeman-shifted resonance frequencies $\omega_m = \omega_0 \pm m\Delta_B$, $m = \pm 1$. Random frequency shifts $\Delta_n$ responsible for inhomogeneous broadening of atomic spectral lines are taken from a centered normal distribution with a variance $\Delta_0^2$.

We model the motion of impurity atoms by oscillations about random equilibrium positions $\vec{r}_n^{(0)}$ at a fixed radial frequency $\Omega$: $r_{n\mu}(t) = r_{n\mu}^{(0)} + A_{n\mu} \cos(\Omega t + \varphi_{n\mu})$, $\mu = x,y,z$, with random, normally distributed uncorrelated amplitudes $A_{n\mu}$,
$\langle A_{n\mu} \rangle = 0$, $\langle A_{n\mu}^2 \rangle = A^2$,  and
uniformly distributed uncorrelated phases $\varphi_{n\mu} \in [0, 2\pi)$. Assuming that all impurity atoms oscillate independently but at the same frequency $\Omega$ is of course an approximation to a real situation where a distribution of oscillation frequencies can be expected. However, such an approximation known as Einstein's model of a solid \cite{einstein1907,hook95,rogers05} is rather accurate for the model of heat capacity of solids at not-too-low temperature. We thus expect it to be sufficient for our purposes as well.


Considering a fixed $\Omega$  allows for finding a stationary periodic solution of Eq.\ (\ref{dynamic}). Indeed, assume that the solution $\bm{\beta}(t_0)$ of Eq.\ (\ref{dynamic}) is known for a given time $t_0$ and that the motion of impurity atoms is slow enough to be neglected on a time interval $\Delta t$ that is sufficiently long compared to the time $2R/c$ needed for light to cross the atomic sample ballistically. Here $c$ is the speed of light in the transparent host medium. Then, up to a time $t_0+\Delta t$ the system of equations Eq.\ (\ref{dynamic}) can be approximated by $d\bm{\beta}(t)/dt  = \hat{B}(t_0) \bm{\beta}(t) + \bm{\gamma}$, where $\bm{\beta}$ is a vector of length $3N$. $\hat{B}(t_0)$ is a $3N \times 3N$ matrix, and $\bm{\gamma}$ is a vector of coefficients, both following from Eq.\ (\ref{dynamic}). The solution of such a system of equations for time $t = t_0 + \Delta t$ can be found using the matrix exponential and integrating factors:
\begin{eqnarray}
\begin{aligned}
\bm{\beta}(t_0 + \Delta t)
&= \hat{M}(t_0, \Delta t) \left[ \bm{\beta}(t_0) + \bm{\gamma}(t_0, \Delta t) \right]
\end{aligned}
\label{solution}
\end{eqnarray}
where
\begin{eqnarray}
\begin{aligned}
\hat{M}(t_0, \Delta t) &= e^{\Delta t \hat{B}(t_0)}
\\
\bm{\gamma}(t_0, \Delta t) &=
\hat{B}(t_0)^{-1} \left[1 - e^{-\Delta t \hat{B}(t_0)} \right]
\bm{\gamma}
\end{aligned}
\label{defm}
\end{eqnarray}

For impurity atoms oscillating around their equilibrium positions at radial frequency $\Omega$ and continuous-wave excitation, the steady-state solution $\bm{\beta}(t)$ is periodic with a period $T = 2\pi/\Omega$. Thus, we only need to find $\bm{\beta}(t)$ on a time interval $[t_0, t_0 + T)$, where $t_0$ is arbitrary. To this end, we split this interval in $n_{\text{max}}$ short subintervals of equal duration $\Delta t$ (typically, $n_{\text{max}} \sim 30$ is sufficient for the analysis presented below) and iterate Eq.\ (\ref{solution}):
\begin{eqnarray}
\begin{aligned}
\bm{\beta}(t_n)
&= \hat{M}(t_{n-1}, \Delta t) \left[ \bm{\beta}(t_{n-1}) +  \bm{\gamma}(t_{n-1}, \Delta t) \right]
\end{aligned}
\label{iteration}
\end{eqnarray}
which provides the solution $\bm{\beta}(t)$ at discrete times $t_n$.

Equation (\ref{iteration}) requires knowing $\bm{\beta}(t_0)$ to start the iterations. We circumvent this problems in two different ways. First, in the steady state we should have $\bm{\beta}(t_0 + T) = \bm{\beta}(t_0)$. Expressing $\bm{\beta}(t_0 + T)$ via $\bm{\beta}(t_0)$ using Eq.\ (\ref{iteration}) and solving the resulting equation yields $\bm{\beta}(t_0)$. Second, we can start with arbitrary $\bm{\beta}(0)$ and iterate Eq.\ (\ref{iteration}) long enough (typically, thousands of iterations) to reach a steady-state regime. The result is independent of $\bm{\beta}(0)$ because of the presence of relaxation ($\Gamma_0 > 0$) in Eq.\ (\ref{dynamic}). Both approaches yield the same results whereas the second one has an additional advantage of explicitly proving the existence and stability of the periodic solution.

The ability to solve Eq.\ (\ref{dynamic}) allows for describing the transport of electromagnetic radiation through the ensemble of oscillating impurity atoms. In particular, the electric field outside the ensemble is found as
\begin{eqnarray}
\begin{aligned}
\vec{E}(\vec{r}, t) &= \vec{E}_0(\vec{r}) - 4\pi d k_0^2
\sum\limits_{n = 1}^{N} \hat{G}\left[
\vec{r}-\vec{r}_n(t) \right]
\hat{d}^{\dagger}\bm{\beta}_{n}(t).
\end{aligned}
\label{field}
\end{eqnarray}
To estimate the transmission coefficient $\langle T_a \rangle$ of an infinitely wide slab, we perform simulations in cylindrical samples of radius $R > 3L$, see the inset of Fig.\ \ref{fig_trans}. Equations (\ref{dynamic}) are solved numerically for an incident plane wave $\vec{E}_0(\vec{r},t) = \vec{u}_0 E_0 \exp(i k z - i\omega t)$ and the transmitted electric field $\vec{E}(\vec{r},t)$ is calculated using Eq.\ (\ref{field}) in a plane located at some distance behind the sample (we use $z = L + 12/k_0$ in the calculations presented below). Typical behavior of ensemble averages of the time-averaged intensity
\begin{eqnarray}
\langle I(\vec{r}) \rangle = \left\langle \frac{\Omega}{2\pi}
\int\limits_0^{2\pi/\Omega}
|\vec{E}(\vec{r},t)|^2 dt \right\rangle
\label{int}
\end{eqnarray}
is illustrated in Fig.\ \ref{fig_trans}. Overall, $\langle I(\vec{r}) \rangle$ exhibits a behavior that could be expected for diffraction of a plane wave on a cylindrical obstacle, with pronounced oscillations at large distances $|x| > R$ from the cylinder axis. At the same time, the transmitted intensity flattens near the axis, in particular for $|x| < R_1 \simeq R/2$, at a level that becomes less and less sensitive to $R$ as $R$ increases. This justifies using a spatial average of $\langle I(\vec{r}) \rangle$ over $\sqrt{x^2 + y^2} < R_1$ to estimate $\langle T_a \rangle$ that would be obtained for an infinitely wide sample. It should be kept in mind that the choice of the precise radius $R_1$ of the disk  over which the spatial averaging is performed is somewhat arbitrary and is dictated by the need to find a trade-off between the accuracy of calculations and the required computer resources. Choosing smaller $R_1$ would yield a result that would be closer to the infinite-$R$ limit but would require averaging over a larger number of independent atomic configurations to get rid of statistical fluctuations. On the contrary, using larger $R_1$ would ensure better averaging but would bring the result further from the desired $R \to \infty$ limit. We do not expect any of our conclusions to depend on the precise choice of $R_1$ as far as $R_1 < R$.

The time- and ensemble-averaged transmission coefficient follows from Eq.\ (\ref{int}):
\begin{eqnarray}
\begin{aligned}
\langle T_a \rangle &= \frac{\Omega}{2 \pi^2 R_1^2 |E_0|^2} \int\limits_0^{2\pi/\Omega} dt \int\limits_{\pi R_1^2} d^2 \vec{r} \langle \left| \vec{E}(\vec{r},t) \right|^2 \rangle
\end{aligned}
\label{intaver}
\end{eqnarray}
where the spatial integration is over a disk of radius $R_1 = R/2$ in a distant plane $z = L + 12/k_0$ and with a center on the $z$ axis, angular brackets denote averaging over random sets of equilibrium positions $\{ \vec{r}_n^{(0)} \}$, amplitudes $\{ A_{n\mu} \}$ and phases $\{ \varphi_{n\mu} \}$ of atomic oscillations
as well as over random frequency shifts $\Delta_n$. The subscript `$a$' of $T_a$ refers to the mode of light incident on the atomic sample. In our calculations, it is a circularly polarized plane wave exciting the transition $|g\rangle \to |e_m\rangle$ under study. Averaging over $\{ \varphi_{n\mu} \}$ reflects the fact that impurity oscillations are due to thermal phonons that do not have long-time coherence. This averaging  suppresses periodic oscillations of $\langle \left| \vec{E}(\vec{r},t) \right|^2 \rangle$ and makes it independent of time, apart from remaining statistical fluctuations.

\section{Signatures of Anderson localization in transmission of a plane wave}
\subsection{Scaling of dimensionless conductance}

Below we analyze the possibility of Anderson localization using several different criteria. The first one is based on the study of the average effective dimensionless conductance of the considered samples. Dimensionless conductance $g$---the inverse of resistance in units of $e^2/h$ for spinless electrons---is a very useful quantity to study Anderson localization. Defined for a cube of side $L \gg \ell$, $g$ increases with $L$ for diffuse scattering, decreases with $L$ when Anderson localization sets in, and becomes independent of $L$ at the localization transition (mobility edge) \cite{abrahams79,kramer93}. This allows for detecting Anderson localization by analyzing the scale dependence of $g$.

For classical and, in particular, electromagnetic waves, samples are usually open at all sides and waves can escape from the sample into the surrounding free space surrounding \cite{sheng06}. This is in contrast to electrons that can only propagate through leads attached to the sample. For this reason, working with cube-shaped samples is not very practical in both experiments and theory: leakage of waves through sample sides reduces $g$ and compromises the scaling analysis. Instead, optical experiments are often conducted in slab-shaped samples that are so wide that can be assumed infinitely extended in the transverse direction \cite{wiersma97,storzer06,sperling13}. This is the geometry that we model in the present work by considering a cylindrical sample shown in the inset of Fig.\ \ref{fig_trans}. Ideally, we would want to have $R \gg L$ but practical limitations in computer memory and calculation time only allow us to work with $R$ up to (3--15)$L$.

For transmission of a plane electromagnetic wave $\propto \exp(i k z)$ through a cylindrical sample shown in the inset of Fig.\ \ref{fig_trans}, the average conductance can be related to the average total transmission coefficient $\langle T_a \rangle$ as $\langle g \rangle = (4/5) N_{\perp} \langle T_a \rangle$, where $N_{\perp} = k^2 (\pi R^2)/2\pi$ is the number of transverse modes and we assume standard boundary conditions for the diffuse energy flux with an extrapolation length $z_0 = 2\ell/3$ \cite{akkermans06,yamilov24}.
More accurate treatment of boundary conditions would only affect the numerical prefactor $4/5$, which is irrelevant for our analysis because we only rely on the scaling of $\langle g \rangle$ with $L$ and not on its value.  On the one hand, $\langle g \rangle$ defined in this way can be made arbitrary large simply by increasing $R$. On the other hand, it always decreases with the sample thickness $L$ (because $\langle T_a \rangle$ decreases with $L$) for any strength of scattering. Such properties make it useless in the context of Anderson localization. However, we can still define a quantity that would have the same scaling properties as the dimensionless conductance of a cube-shaped disordered conductor in the case of electron transport by replacing $R$ by $L$ in the expression for $N_{\perp}$:
\begin{equation}
\langle {\tilde g} \rangle = \frac{2}{5} (k L)^2 \langle T_a \rangle
\label{gdef}
\end{equation}
${\tilde g}$ is not equal to the sum of transmission coefficients corresponding to all possible incident modes (plane waves in the limit of $R \to \infty$); this equality holds for $g = \sum_a T_a \ne {\tilde g}$. However, $\langle {\tilde g} \rangle$ has the desired scaling properties: it grows with $L$ for diffuse scattering when $\langle T_a \rangle \propto 1/L$, decreases with $L$ when Anderson localization sets in because $\langle T_a \rangle \propto \exp(-L/\xi)$ in this case, and becomes independent of $L$ at the mobility edge where $\langle T_a \rangle \propto 1/L^2$ \cite{anderson85,sheng06}.
Note that our definition of $\langle {\tilde g} \rangle$ only makes sense for $R, L > \ell$, which is well obeyed in the calculations reported below because $k_0\ell \lesssim 1$ under considered conditions (see Appendix \ref{sec:mfp}).

Substituting $L$ for $R$ in the expression of $\langle g \rangle$ is not just a mathematical trick but is physically motivated by the isotropy of light scattering on average. Indeed, light that traverses a sample of thickness $L$ should also explore a volume of transverse extent $R \sim L$. Features of the sample beyond $x^2 + y^2 \sim L^2$ should not be relevant for measurements around the $z$ axis. Thus, the relevant transverse extent of the sample of thickness $L$ is of the order of $L$ as well.
\begin{figure}[t]
\hspace*{6.5mm}
\includegraphics[width=1.0\columnwidth]{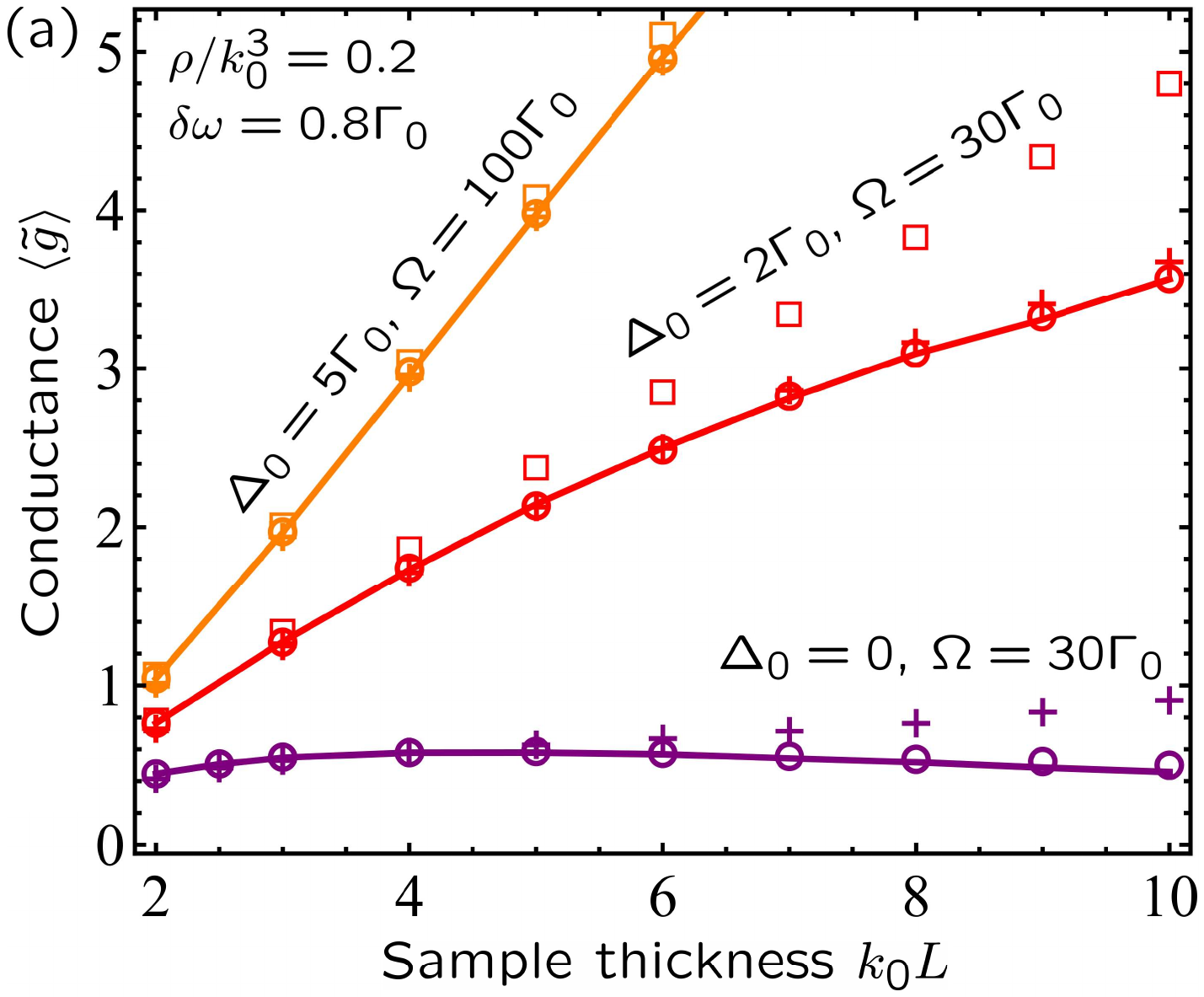}\\
\vspace*{-2mm}
\hspace*{4mm} \includegraphics[width=1.0\columnwidth]{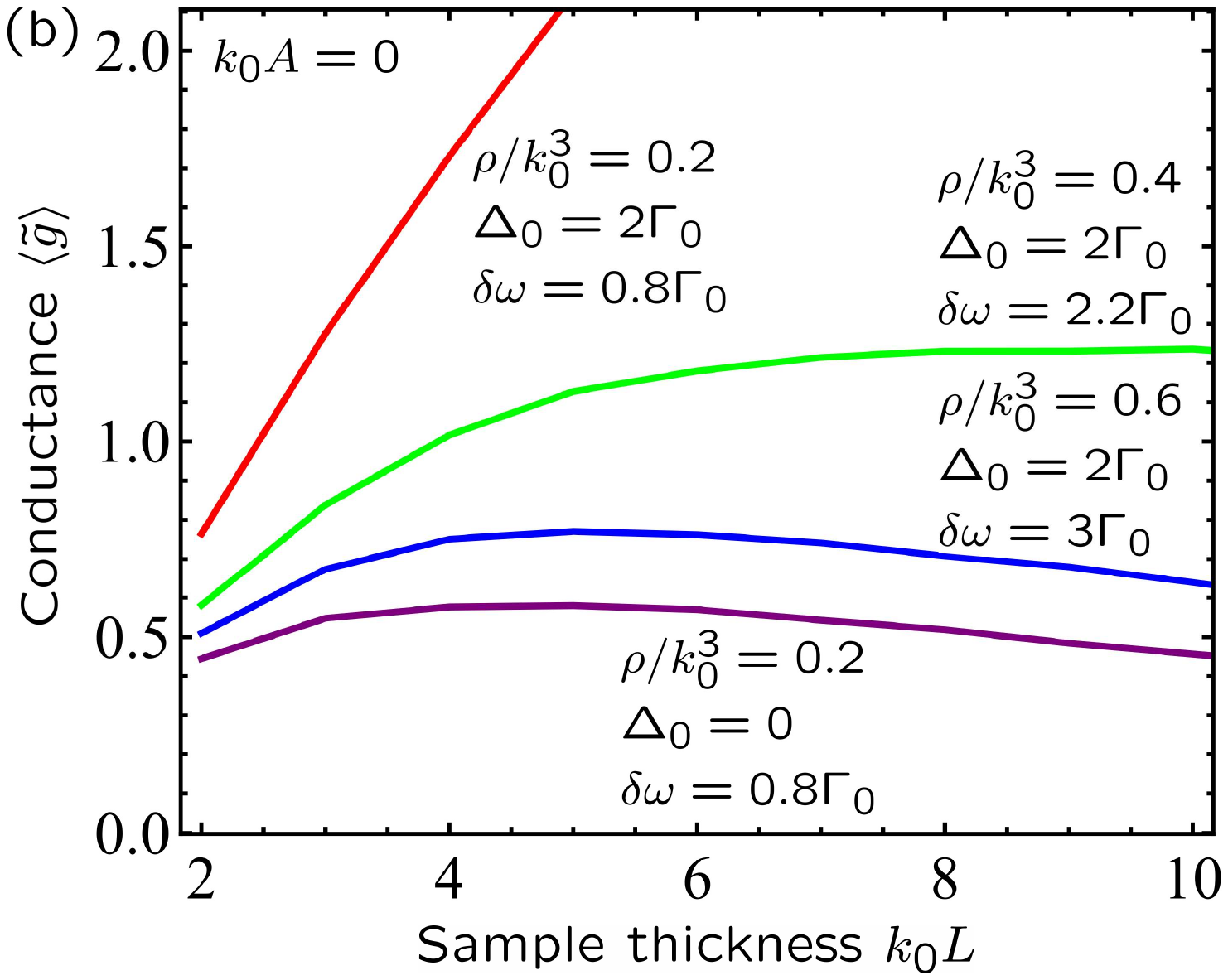}
\vspace*{-7mm}
\caption{\label{fig_g}
(a) Impact of oscillations of impurity atoms on the dimensionless conductance $\langle {\tilde g} \rangle$ of a cylinder-shaped sample of radius $k_0 R = 30$ and thickness $L$ depicted in the inset of Fig.\ \ref{fig_trans}. Symbols show results of full dynamic calculations accounting for fast oscillations of impurity atoms around random equilibrium positions with root-mean-square amplitude $A$ and  frequency $\Omega$
as well as random frequency shifts of impurity resonant frequencies (variance $\Delta_0^2$).
Open circles correspond to $k_0 A = 0.02$, crosses to $k_0 A = 0.1$, open squares to $k_0 A = 0.3$. Solid lines are obtained for $k_0 A = 0$.
(b) Impact of inhomogeneous broadening of atomic transition lines on dimensionless conductance $\langle {\tilde g} \rangle$ of the same sample as in (a) for $k_0 A = 0$.
}
\end{figure}

Figure \ref{fig_g} shows dependencies of $\langle {\tilde g} \rangle$ on $L$  for different
sets of parameters and leads us to several important conclusions. First, Fig.\ \ref{fig_g}(a) illustrates that the impact of atomic oscillations is negligible in both diffusion
(the upper orange line and symbols) and localization
(the lower purple line and symbols) regimes of propagation
as far as $k_0 A \ll 1$. This can be understood by analogy with the so-called Dicke effect that consists in narrowing of atomic spectral lines in the presence of inter-atomic collisions:
when the mean free path of atoms between two collisions becomes much shorter than the wavelength of light, the atomic line width becomes insensitive to the Doppler effect \cite{dicke53}. In our case, oscillations of atoms replace their random displacements due to collisions, and the condition of insensitivity of our results to oscillations becomes $k_0 A \ll 1$.
This condition is readily satisfied in a typical solid where $A \lesssim 1$\AA\,  and $k_0 A \lesssim 10^{-3}$ for visible light, allowing us to
neglect atomic oscillations
in the following. However, atomic oscillations start to play a role and can conceal Anderson localization already for $k_0 A \gtrsim 0.1$ as can be seen from the results shown by crosses and open squares in Fig.\ \ref{fig_g}(a).

Next, Fig.\ \ref{fig_g}(b) shows that the destructive effect of inhomogeneous broadening on localization can be mitigated by increasing the atomic number density $\rho$ and adjusting the detuning $\delta\omega$. In particular, whereas the lower (purple) line obtained in the absence of inhomogeneous broadening ($\Delta_0 = 0$) decays with $k_0 L$ for $k_0 L \gtrsim 4$ indicating Anderson localization, the upper (red) line obtained for exactly the same parameters except for $\Delta_0 = 2\Gamma_0$, grows with $k_0 L$ signaling that Anderson localization is suppressed by the inhomogeneous broadening.  However, when we keep $\Delta_0$ constant and increase the atomic number density $\rho$ adjusting the frequency detuning $\delta\omega$, the curve $\langle {\tilde g}(k_0 L) \rangle$ bends down and approaches the one in the absence of broadening [see the two intermediate green and blue curves in Fig.\ \ref{fig_g}(b)]. Whereas increasing $\rho$ may be difficult for cold-atom systems, it is not necessarily the same for solid-state samples with embedded impurities. This gives hope for observation of Anderson localization of light in such samples despite the inhomogeneous broadening phenomenon.

\subsection{Spatial profile of the average atomic excitation}

\begin{figure}[t]
\hspace*{4mm}
\includegraphics[width=1.0\columnwidth]{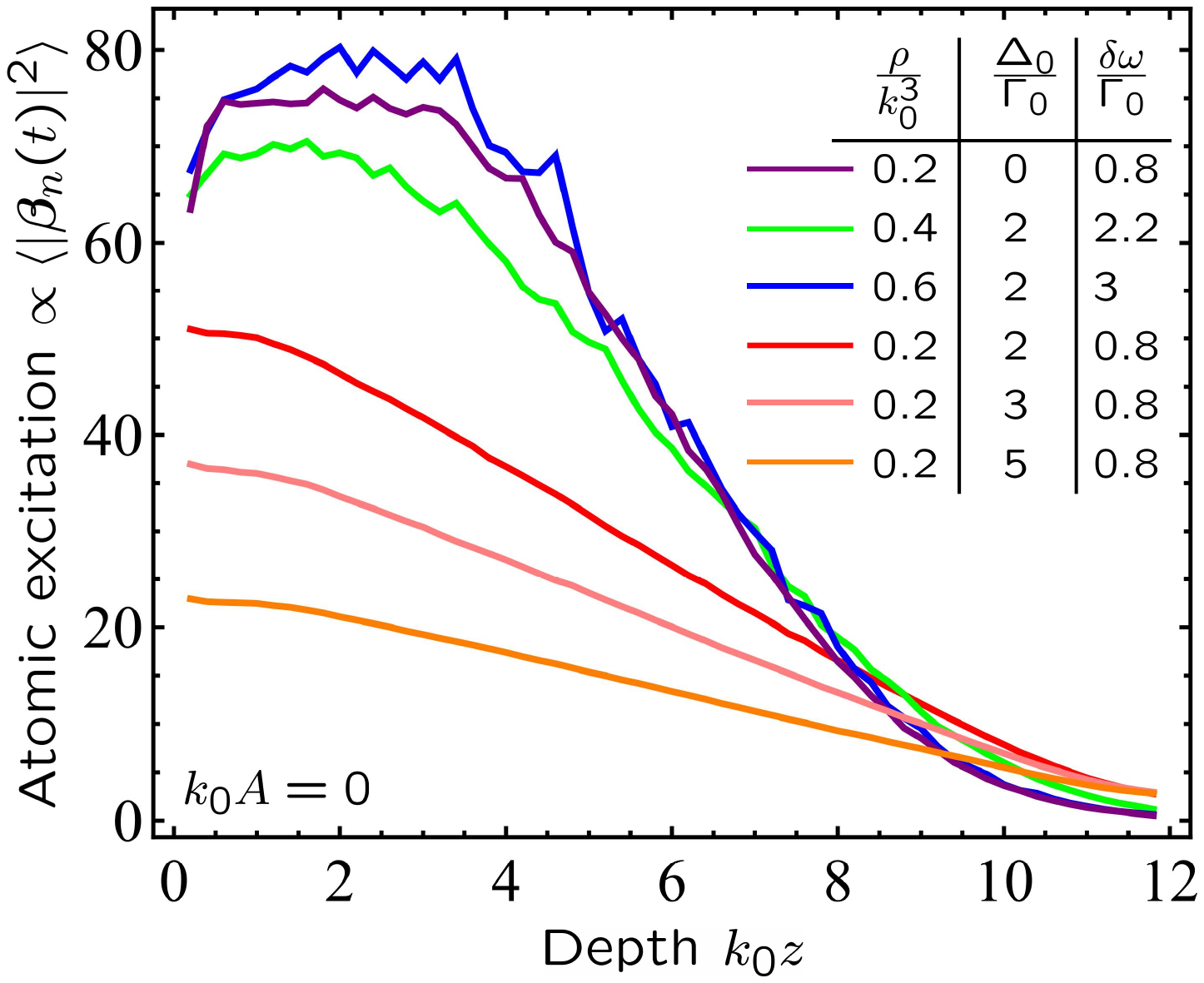}
\vspace{-9mm}
\caption{\label{fig_inside}
Average profiles of atomic excitation inside the
cylindrical sample
of radius $k_0 R = 25$ and
thickness $k_0L = 12$
(see the inset of Fig.\ \ref{fig_trans}),
for parameters listed in the inset.}
\end{figure}

In addition to the dependence of dimensionless conductance on sample thickness, a signature of Anderson localization can be found in the
depth profile
of the average atomic excitation $\langle |\bm{\beta}_{n}(t)|^2 \rangle$, with the averaging also over time $t$,
on the depth $z$ inside the sample,
see Fig.\ \ref{fig_inside}. All curves in Fig.\ \ref{fig_inside} correspond to atomic densities $\rho$ for which Anderson localization is expected at $\delta\omega/\Gamma_0 = 0.8$ for $\Delta_0 = 0$ \cite{skip15prl,skip18prl}. We thus expect a step-like (steep in the middle of the sample and flattened towards $z=0$ and $z=L$) profile of $\langle |\bm{\beta}_{n}(t)|^2 \rangle$
illustrated by the purple curve in Fig.\ \ref{fig_inside} \cite{skip19prl}. However, it is clear that the three lower curves obtained for the same $\rho$ and $\delta\omega$ as the purple curve but in the presence of inhomogeneous broadening $\Delta_0 \geq 2$, exhibit a roughly linear decay with $z$ characteristic of photon diffusion \cite{skip19prl,tiggelen21}. Thus, the inhomogeneous broadening suppresses Anderson localization. Nevertheless, similarly to what has been discussed in connection with Fig.\ \ref{fig_g}, this suppression can be mitigated and even fully canceled by increasing $\rho$ and adjusting $\delta\omega$, as clearly witnessed by the green and blue curves in Fig.\ \ref{fig_inside} that exhibit the shape expected for Anderson localization.

\subsection{Fluctuations of intensity}
\label{sec:fluct}

\begin{figure}[t]
\hspace*{5mm}
\includegraphics[width=\columnwidth]{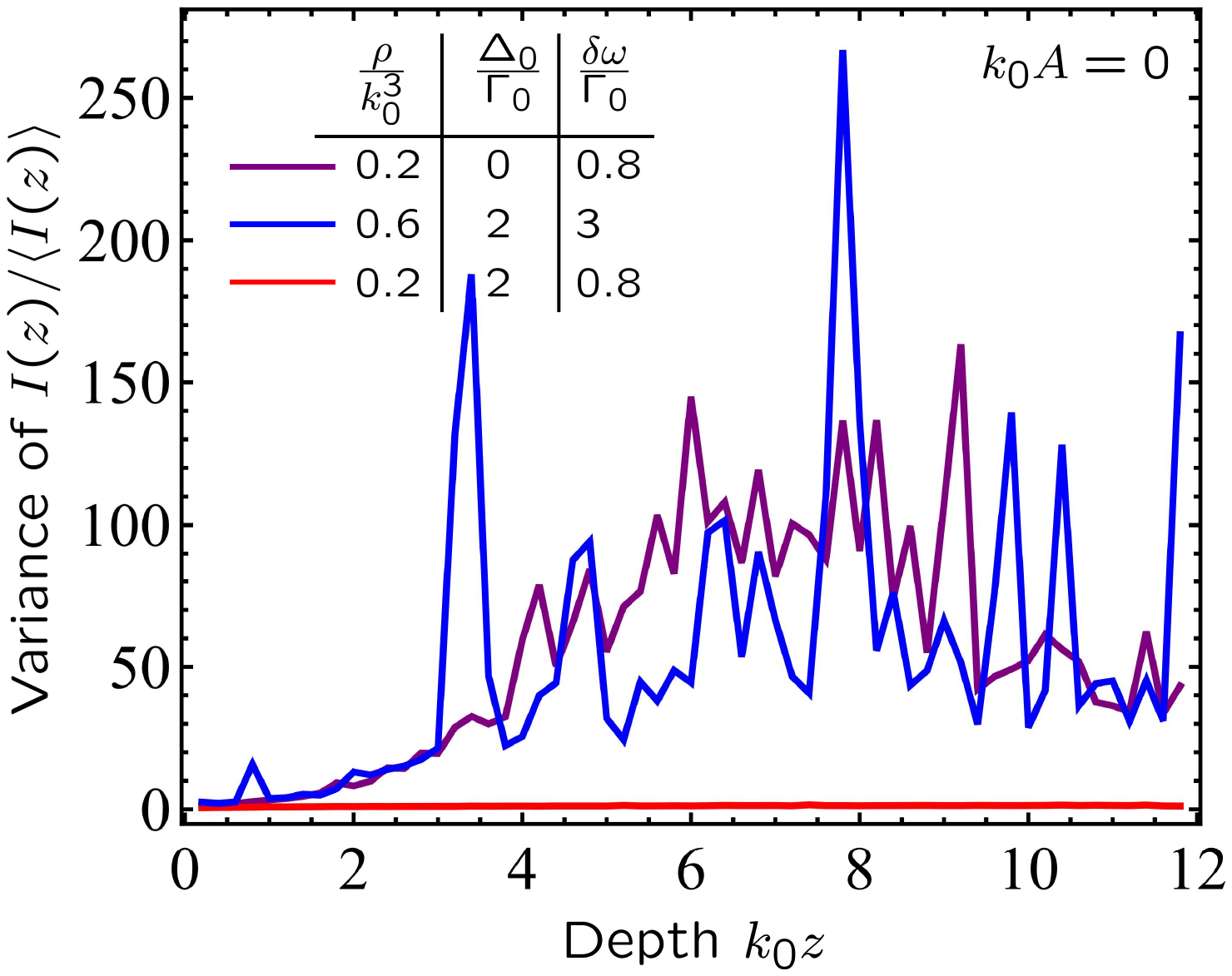}
\vspace*{-7mm}
\caption{\label{fig_var}
Normalized variance of transverse position-averaged
intensity as a function of depth $z$ inside the sample
for $k_0 A = 0$.
Ensemble averaging is performed over
a large number of random atomic configurations ranging from $3 \times 10^4$ for $\rho/k_0^3 = 0.6$ to $3 \times 10^5$ for $\rho/k_0^3 = 0.2$. Parameters of calculations are the same as for the curves of the same colors in Figs.\ \ref{fig_g} and \ref{fig_inside}.}
\end{figure}

Another confirmation of Anderson localization of light in the presence of inhomogeneous broadening of atomic spectral lines can be obtained by analyzing fluctuations of intensity $I(\vec{r},t) = |\vec{E}(\vec{r},t)|^2$. In our model, the probability of excitation of an atom located at a point $\vec{r}_n$ is proportional to the intensity of light at this point: $|\bm{\beta}(\vec{r},t)|^2 \propto I(\vec{r},t)$. Hence, the normalized variance of intensity inside the atomic cloud is equal to the normalized variance of atomic excitation probability.
Assuming $k_0 A \ll 1$, we neglect atomic oscillations and average $I(\vec{r},t)$ over the crosssection of the cylindrical sample to obtain
\begin{eqnarray}
I(z) = \frac{1}{\pi R_1^2} \int_{\pi R_1^2} d^2 \vec{r}_{\perp} I (\vec{r})
\label{intz}
\end{eqnarray}
where $\vec{r}_{\perp} = \{x, y \}$.
The normalized variance of this quantity
$\text{var} I (z)/\langle I (z) \rangle^2$ is shown in Fig.\ \ref{fig_var}.

Figure \ref{fig_var} clearly shows that fluctuations of $I(z)$ are much stronger for the values of parameters that we associate with Anderson localization in Figs.\ \ref{fig_g} and \ref{fig_inside}  (purple and blue curves) as compared to the values corresponding to diffuse scattering (red curve). Actually, the fluctuations are so large that we did not manage to obtain smooth curves for $\text{var} I (z)/\langle I (z) \rangle^2$ even after averaging over several hundreds of thousands of independent atomic configurations. This provides an additional proof of Anderson localization that is known to be associated with strong intensity fluctuations \cite{chabanov00,skip19prl,cottier19,yamilov23}.

\section{Quasimode localization analysis}
\label{sec:quaismodes}

To gain a systematic understanding of the impact of inhomogeneous broadening on Anderson localization, we analyze the statistical properties of complex eigenenergies $E_n$ and right eigenvectors (also called quasimodes) $\bm{\psi}_n$ of the effective Hamiltonian (\ref{heff}). We generate independent spatial configurations of $N$ atoms randomly distributed inside a sphere of radius ${\cal R}$ at a number density $\rho = N/V$ with $V = 4\pi {\cal R}^3/3$. For each atom, a random shift $\Delta_n$ of its resonance frequency $\omega_0$ is sampled from a centered normal distribution with variance $\Delta_0^2$. For each atomic configuration with a corresponding set of $\Delta_n$, complex eigenvalues $E_n = -2(\omega_n-\omega_0)/\Gamma_0 + i\Gamma_n/\Gamma_0$
and eigenvectors $\bm{\psi}_n$
of the  Hamiltonian (\ref{heff}) are computed numerically and ordered according to the real part of $E_n$.

\subsection{Inverse participation ratio}
\label{sec:ipr}

\begin{figure*}[t]
\includegraphics[width=0.83\textwidth]{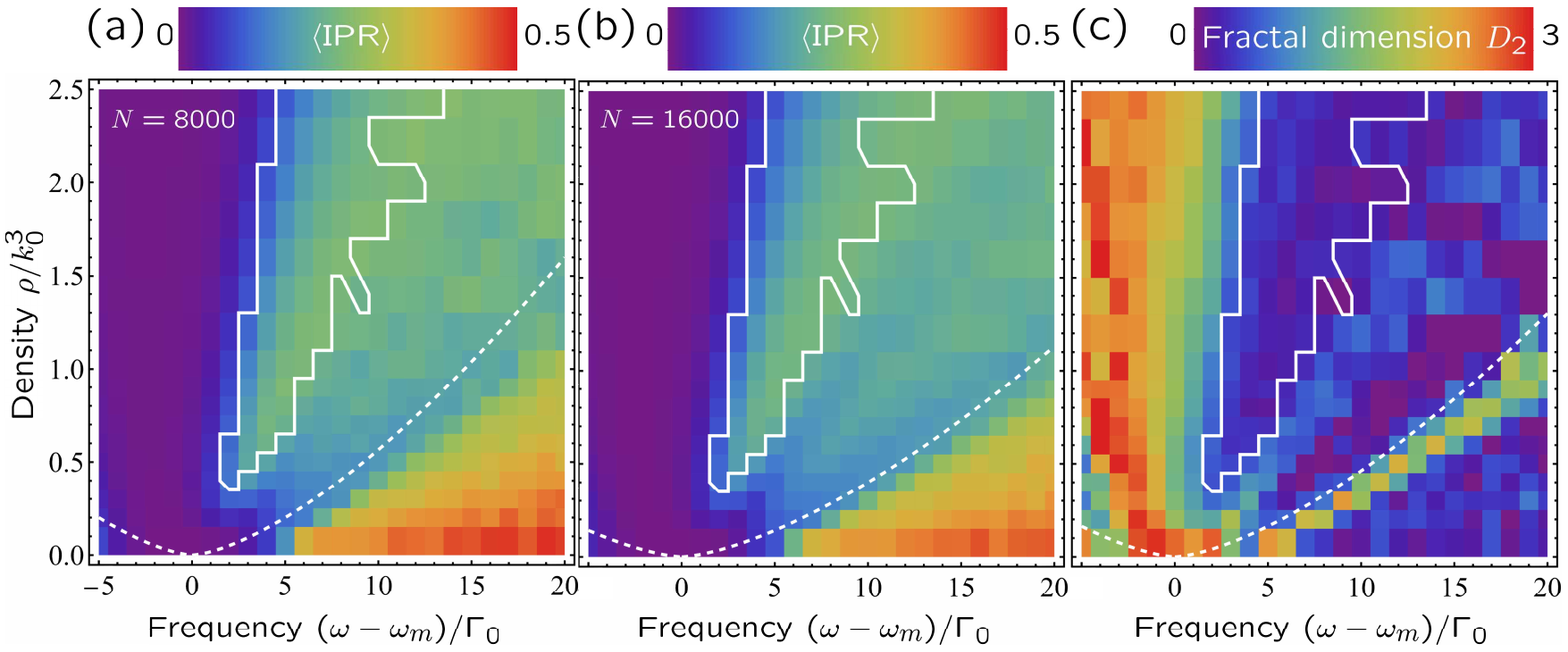}
\vspace*{-4.5cm}
\caption{\label{fig_ipr}
Color-scale plot of IPR for $\Delta_0/\Gamma_0 = 2$, $N = 8000$ (a) and 16000 (b) averaged over quasimodes with different decay rates $\Gamma_n$ and over 30 and 25 independent configurations of $N$ atoms in a sphere of radius ${\cal R}$, respectively, as a function of quasimode frequency $\omega$. (c) shows the fractal dimension defined by $\langle \text{IPR} \rangle \propto {\cal R}^{-D_2}$. The color scale is capped between physically meaningful values 0 and 3 whereas the calculated values of $D_2$ vary between $-0.69$ and $3.18$, which we attribute to insufficient averaging. Solid white contours show the mobility edge $\beta = 0$. Dashed white lines are Eq.\ (\ref{manyvs2}), with $N = 12000$ for the panel (c).}
\end{figure*}

We quantify the spatial localization of eigenvectors $\bm{\psi}_n$ by their inverse participation ratios (IPR)
\begin{eqnarray}
\text{IPR}_n = \frac{\sum_{m=1}^N \left( \sum_{\mu = 1}^3 |\psi_n^{(\mu)}(\vec{r}_m)|^2 \right)^2}{\left( \sum_{m=1}^N \sum_{\mu = 1}^3 |\psi_n^{(\mu)}(\vec{r}_m)|^2 \right)^2}
\label{eqipr}
\end{eqnarray}
where $\psi_n^{(\mu)}(\vec{r}_m)$ is the value of the $\mu$th polarization component of $\bm{\psi}_n$ at the atom $m$. In general, IPR varies from $1/N$ for eigenvectors extended over all $N$ atoms to 1 for eigenvectors localized on a single atom. In disordered atomic ensembles considered here, the most localized eigenvectors extend over a pair of closely located atoms and correspond to so-called proximity resonances \cite{heller96,rusek00} for which $\text{IPR} \simeq 0.5$. We show a color-scale plot of average IPR as a function of frequency $\omega$ and atomic number density $\rho$ in Figs.\ \ref{fig_ipr}(a) and (b) for a representative strength of inhomogeneous broadening $\Delta_0/\Gamma_0 = 2$ and two different number of atoms $N = 8000$ and 16000. Eigenvectors with $\langle \text{IPR} \rangle \simeq 0.5$ present at all densities for large frequency shifts are due to eigenvectors localized on pairs of closely located atoms. To isolate the contribution of the latter, we refer to the previous work \cite{skipetrov11,goetschy11a} that has demonstrated that the complex eigenvalues $E_n$ of the matrix $\hat{H}_{\text{eff}}$ corresponding to many-atom states are mainly concentrated within a circular region of radius $\gamma/2$ on the complex plane, where $\gamma = 9N/8(k_0 {\cal R})^2$. Two-atom states give rise to ``branches'' of eigenvalues outside this region. These results were obtained for the scalar model of light scattering but remain relevant for the full vector model with a strong magnetic field that lifts the degeneracy of atomic transitions corresponding to different magnetic quantum numbers $m = 0, \pm 1$. Thus, a boundary between many- and two-atom states is roughly determined by a condition $2|\omega-\omega_m|/\Gamma_0 = \gamma/2$ leading to
\begin{eqnarray}
\frac{\rho}{k_0^3} = \frac{1.6}{\sqrt{N}} \left| \frac{\omega-\omega_m}{\Gamma_0} \right|^{3/2}
\label{manyvs2}
\end{eqnarray}
This equation is shown by dashed lines in Fig.\ \ref{fig_ipr}. Parts of the plots below dashed lines are dominated by two-atom states and should be discarded in the analysis of Anderson localization that is responsible for the behavior above the dashed lines. We clearly see that quasimodes with substantial IPR arise at sufficiently high densities (for $\rho/k_0^3 \gtrsim 0.3$ for the considered $\Delta_0$) despite the inhomogeneous broadening. The range of frequencies at which $\langle  \text{IPR} \rangle$ is large widens with density.

\subsection{Fractal dimension}
\label{sec:d2}

An additional evidence of Anderson localization can be obtained by computing the fractal dimension $D_2$ of quasimodes defined via a scaling relation $\langle \text{IPR} \rangle \propto {\cal R}^{-D_2}$. Figure \ref{fig_ipr}(c) shows $D_2$ estimated from $\langle \text{IPR} \rangle$ for $N = 8000$ and 16000 in Figs.\ \ref{fig_ipr}(a) and (b), respectively. $D_2 \simeq 3$ for those regions of parameters where $\langle \text{IPR} \rangle$ is low and quasimodes are extended, whereas $D_2 \simeq 0$ when $\langle \text{IPR} \rangle$ is large and quasimodes are localized. This is the behavior expected from the general theory of Anderson localization \cite{evers08}.

\begin{figure*}[t]
\includegraphics[width=0.83\textwidth]{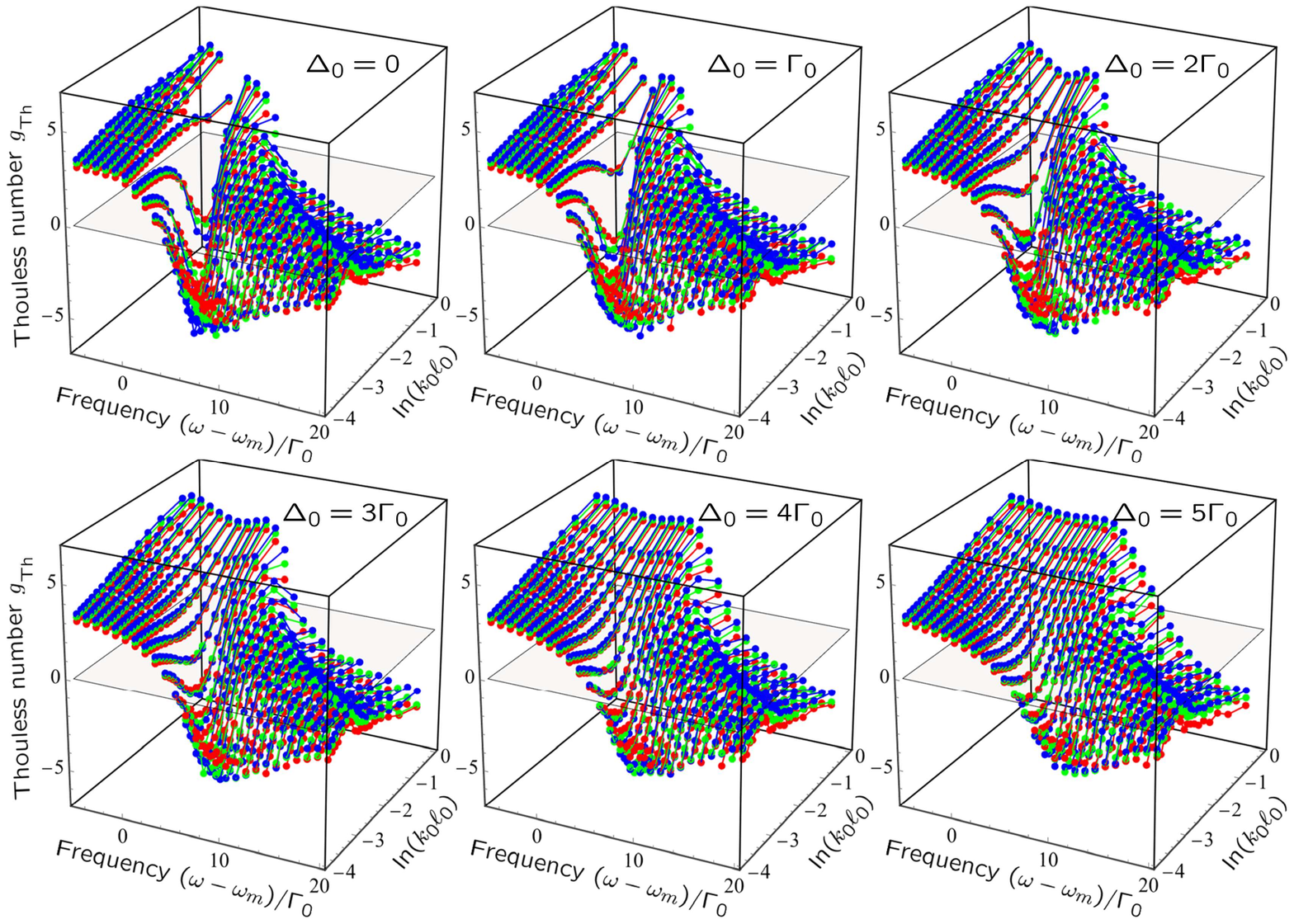}
\vspace*{-3mm}
\caption{\label{fig_g3d}
Thouless number $g_{\text{Th}}$ as a function of the bare Ioffe-Regel parameter $k_0\ell_0 = k_0^3/6\pi\rho$ and frequency detuning $\delta\omega = \omega-\omega_m$, for six different values of inhomogeneous broadening $\Delta_0$.
Data obtained for different total numbers of atoms are shown in different colors: $N = 8000$ (red), 12000 (green), 16000 (blue). The explored range of $k_0\ell_0$ corresponds to the range $\rho/k_0^3 = 0.1$--2.5 of the atomic number density.}
\end{figure*}

\begin{figure*}
\includegraphics[width=0.83\textwidth]{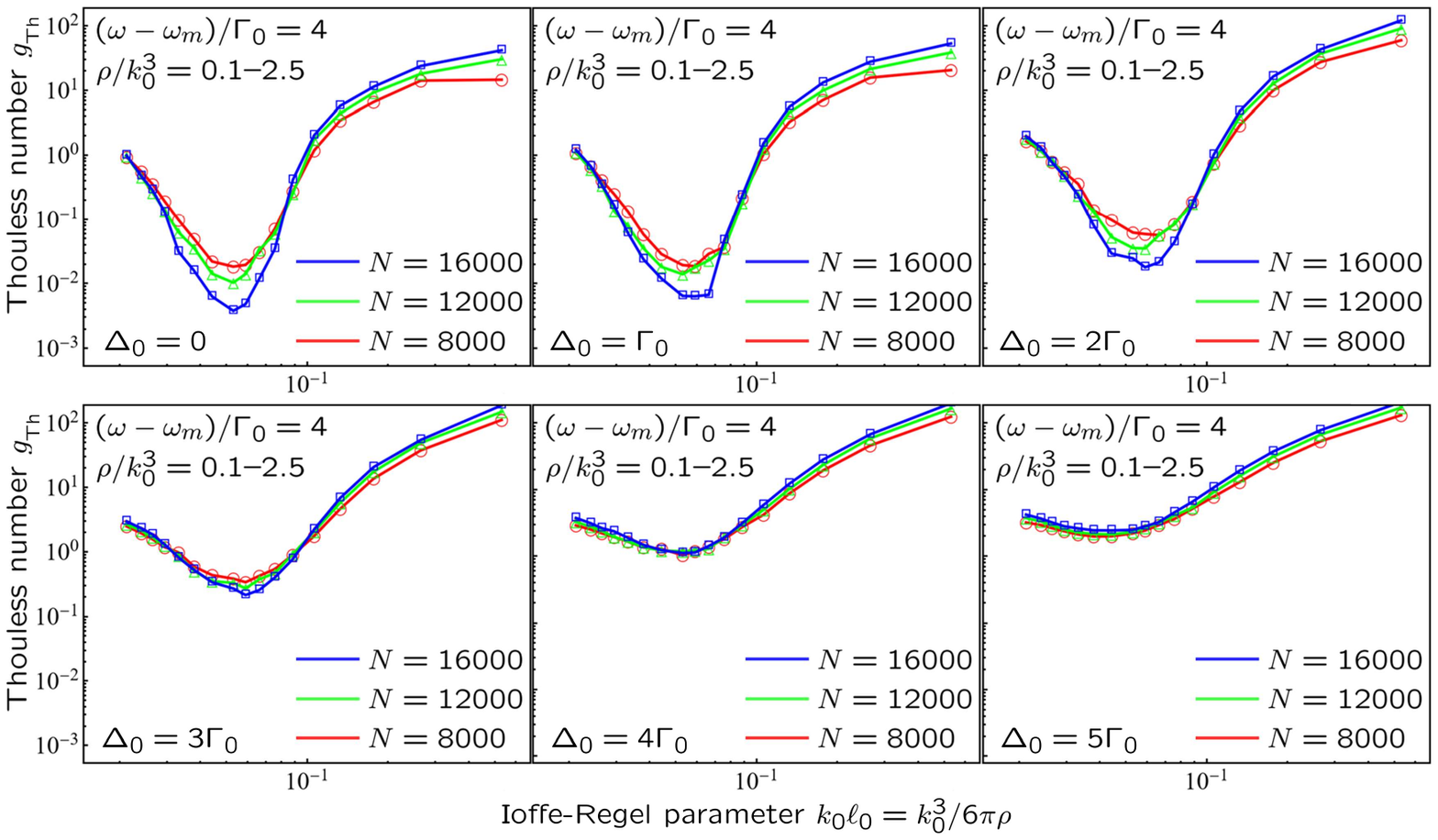}
\vspace*{-2.0cm}
\caption{\label{fig_g2d}
A cut of Fig.\ \ref{fig_g3d} at $\delta\omega = \omega-\omega_m = 4\Gamma_0$, intended to demonstrate the transition from Anderson localization for $\Delta_0 < 4$ (curves corresponding to different $N$ cross in points determining mobility edges) to diffuse transport for $\Delta_0 > 4$ (no crossings).}
\end{figure*}

\begin{figure*}
\includegraphics[width=0.83\textwidth]{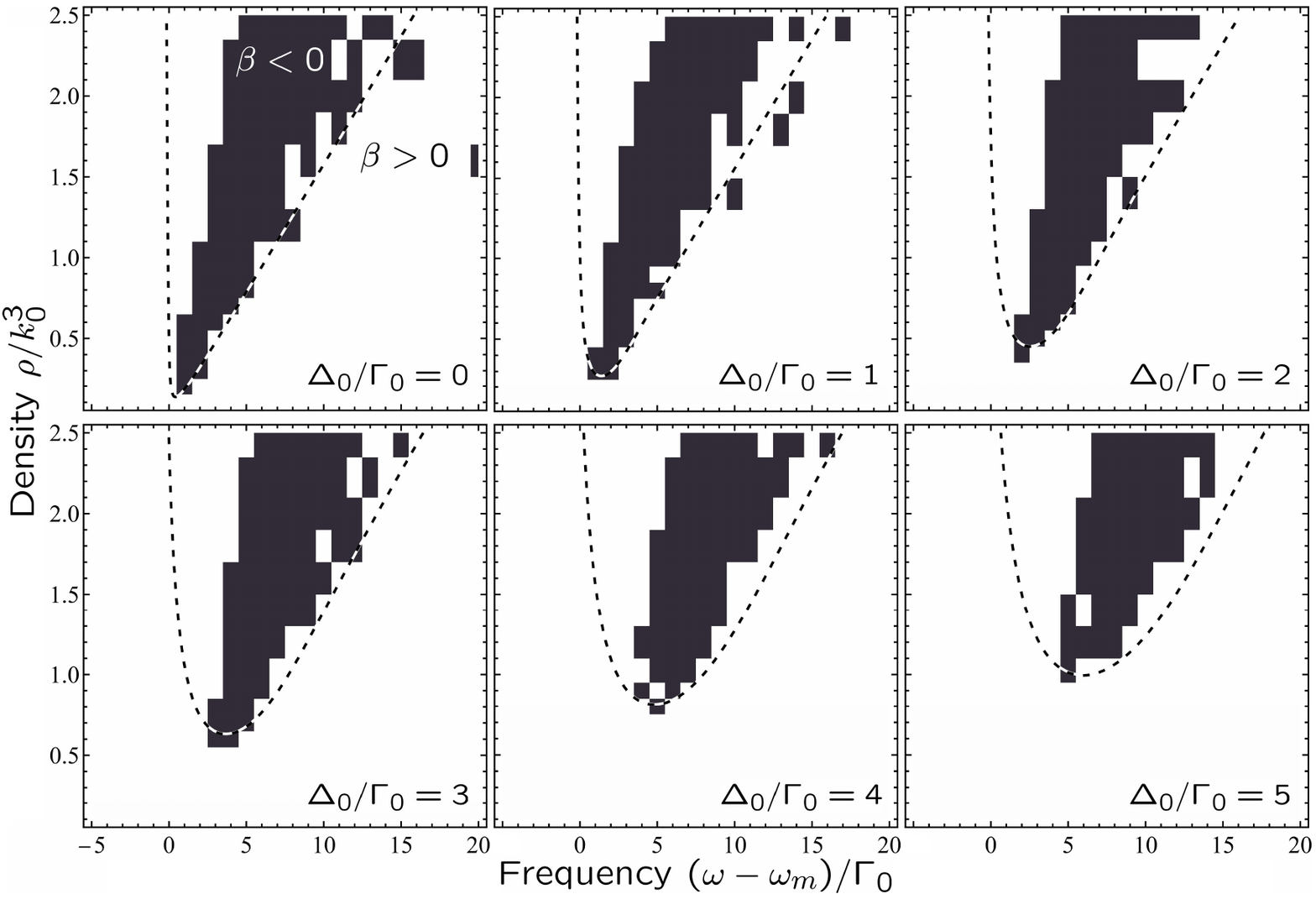}
\vspace*{-5mm}
\caption{\label{fig_phase}
Localization phase diagram for light in a large ensemble of $N \gg 1$ resonant scattering centers embedded in a solid matrix for six values of $\Delta_0/\Gamma_0$. $\beta$-function (approximated by a finite difference from numerical results for $N = 8$ and $16 \times 10^3$) is negative and eigenstates are localized in the region of the phase plane shown in black, whereas $\beta > 0$ and states are extended in the rest of the plane. The dashed line shows the mobility edge $\beta = 0$ predicted by the approximate analytic theory.}
\end{figure*}

\begin{figure}[t]\
\hspace*{4mm} \includegraphics[width=\columnwidth]{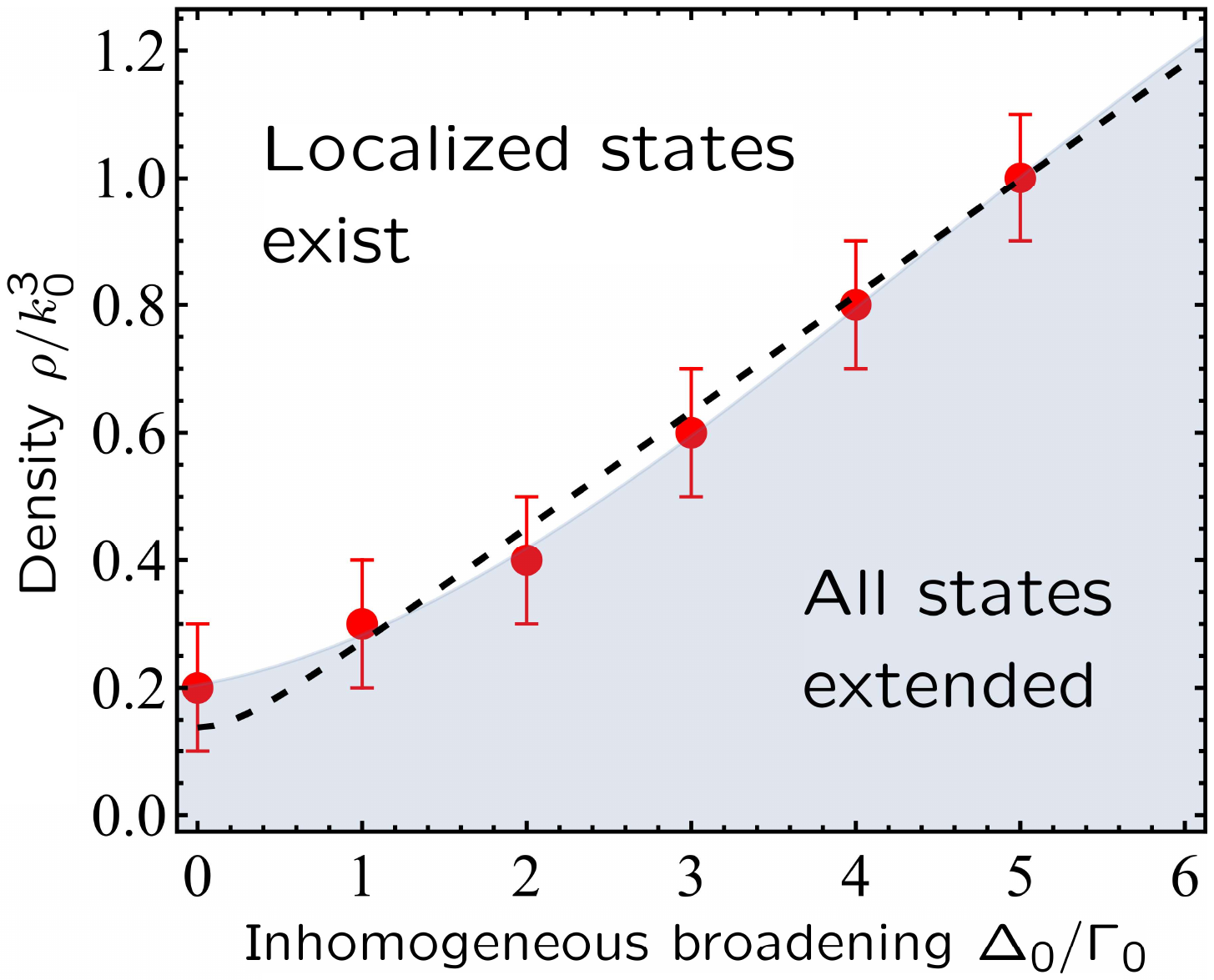}
\vspace*{-7mm}
\caption{\label{fig_min}
Minimum number density of impurity atoms $\rho$ required for collective atomic states to be spatially localized in some frequency band, as a function of inhomogeneous broadening $\Delta_0$ (red disks with error bars). Grey shading shows the region of the parameter space in which all states are extended. Dashed line is the prediction of the approximate analytic theory described in Appendix \ref{approx}.
}
\end{figure}

\subsection{Scaling analysis of Thouless conductance}
\label{sec:scaling}

Another way to prove Anderson localization and distinguish quasimodes that are localized due to disorder in atomic positions from proximity resonances, is to perform a scaling analysis of the so-called Thouless conductance or Thouless number. We compute mean values of $1/\Gamma_n$ and of $\omega_n - \omega_{n-1}$ in each frequency interval of width $\Gamma_0$ from $\delta\omega = \omega-\omega_m = -5\Gamma_0$ to $20\Gamma_0$.
Averaging is also performed over a large number of independent atomic configurations in space and over $\{ \Delta_n \}$. Typically, we average over 50, 35 and 25 configurations for $N = 8000$, 12000 and 16000, respectively. Thouless number is defined as
$g_{\text{Th}}(\omega) = \langle 1/\Gamma_n \rangle_{\omega}^{-1}/\langle \omega_n-\omega_{n-1} \rangle_{\omega}$, where the notation $\langle \cdots \rangle_{\omega}$ highlights the additional averaging over all $\omega_n$ inside an interval of width $\Gamma_0$ around the frequency $\omega$.

Figure\ \ref{fig_g3d} shows $g_{\text{Th}}$ as a function of $k_0 \ell_0 = k_0^3/6 \pi \rho$ and $\delta\omega/\Gamma_0$, for six different values of $\Delta_0$. It is clear that the drop of $g_{\text{Th}}$ around $\delta\omega/\Gamma_0 = 4$ at $\Delta_0 = 0$, shifts to larger $\delta\omega$ and becomes less pronounced with increasing $\Delta_0$. The impact of inhomogeneous broadening on Thouless number at a given frequency is demonstrated in Fig.\ \ref{fig_g2d} for $\delta\omega/\Gamma_0 = 4$. When $\Delta_0 < 4$, localization transitions take place at values of $k_0 \ell_0$ where curves corresponding to different $N$ cross. These transitions are suppressed for $\Delta_0 > 4$, leading to $g_{\text{Th}}$ being a growing function of $N$ at all densities $\rho$.

The $\beta$-function $\beta(g_{\text{Th}}) = \partial \ln g_{\text{Th}}/\partial \ln(k_0 {\cal R})$ can be estimated using a finite-difference approximation for the derivative. Scaling of $g_{\text{Th}}$ with ${\cal R}$ is the same as the scaling of $\langle {\tilde g} \rangle$ with $L$. Thus, the sign of $\beta$ is different for extended ($\beta > 0$) and localized ($\beta < 0$) quasimodes, with $\beta = 0$ defining the mobility edge. The latter is shown by a white solid line in Fig.\ \ref{fig_ipr} and it delimits the left boundary of the region of high $\langle \text{IPR} \rangle$ quite precisely. The right, high-frequency part of the contour $\beta = 0$ is more difficult to identify with a change of  $\langle \text{IPR} \rangle$ in Figs.\ \ref{fig_ipr}(a) and (b) because of two-atom states that start to contribute. However, the fractal dimension $D_2$ in Fig.\ \ref{fig_ipr}(c) clearly exhibits an increase outside the contour $\beta = 0$ all along it. The presented scaling analysis is thus consistent with the analysis of $\langle \text{IPR} \rangle$ and confirms Anderson localization in the considered model. The regions of the parameter space where $\beta < 0$ and quasimodes are spatially localized, are shown in black in Fig.\ \ref{fig_phase} for $\Delta_0/\Gamma_0 = 0$--5.
For $\Delta_0 = 0$, Anderson localization is expected in a certain range of frequencies only when $\rho/k_0^3 \gtrsim 0.1$ \cite{skip18prl}. In agreement with our previous discussion, increasing $\Delta_0$ leads to a requirement of higher density
to reach localization but does not fully suppress it. An additional illustration of this is provided by Fig.\ \ref{fig_min} that shows the minimum density at which Anderson localization takes place in our model at any frequency, as a function of $\Delta_0$. The minimum density increases with $\Delta_0$ but remains finite.

A simple interpretation of results shown in Fig.\ \ref{fig_phase} can be obtained by averaging the self-energy $\Sigma(\omega)$ obtained in the independent scattering approximation (ISA) \cite{sheng06} over the normal distribution of resonance frequencies of individual atoms, see Appendix \ref{approx}. The variance of the distribution is assumed to be equal to a sum of $\Delta_0^2$ due to the random local electric fields and a term $\sim(\Gamma_0 \rho/k_0^3)^2$ due to dipole-dipole interactions between nearby atoms. This turns out to be a simple yet efficient way of taking into account the two effects otherwise neglected by ISA. We define the effective wave number $k_{\text{eff}} = \text{Re}\sqrt{k_0^2 - \langle \Sigma(\omega) \rangle}$ and the scattering mean free path  $\ell = 1/2\text{Im}\sqrt{k_0^2 - \langle \Sigma(\omega) \rangle}$. The Ioffe-Regel criterion of localization yields a simple condition for the mobility edge : $k_{\text{eff}} \ell = (k_{\text{eff}} \ell)_c \sim 1$ \cite{ioffe60,sheng06,skip18prb}. Dashed lines in Fig.\ \ref{fig_phase} show contour plots of this equation for $(k_{\text{eff}} \ell)_c = 0.6$.
They capture the main tendency exhibited by the numerical results although
they are clearly not exact. An even better agreement is obtained for the minimum density required to reach localization in Fig.\ \ref{fig_min}.

\section{Conclusion}
In conclusion, we demonstrate that transparent solids with impurity atoms or ions pinned
at random positions are promising materials for reaching Anderson localization of light in 3D. On the one hand, the detrimental effect of longitudinal optical fields \cite{skip14prl,tiggelen21} can be mitigated in these materials by an external magnetic field, in the same way as in cold-atom systems \cite{skip15prl,skip18prl}, which gives them an advantage over suspensions or powders of dielectric particles \cite{skip16njp,sperling16}. On the other hand,
these materials
do not suffer from strong losses characteristic of metallic structures proposed as candidates for observation of Anderson localization of light \cite{genack91,yamilov23,yamilov24}. The main result of this work is to show that the difficulties specific for solids with impurity atoms---the oscillations of impurity atoms about their equilibrium positions and the inhomogeneous broadening of their spectra due to random local electric fields---are not critical and should not impede observation of Anderson localization of light. Therefore, we believe that it would be worthwhile to put some effort in experimenting with such materials at high number densities of impurity atoms $\rho/k_0^3 \gtrsim 0.2$.

It should be kept in mind that in our analysis, we use a model of impurity atoms oscillating with the same frequency $\Omega$, which is, of course, a simplification with respect to the real situation. Although we believe that this simplification is not essential for our main conclusions to hold, it would be interesting to extend our analysis to non-monochromatic thermal oscillations of atoms. Analyzing such a realistic situation is, however, a much more tedious task that does not allow for relying on the steady-state solutions. Our expectation that considering a single frequency $\Omega$ is sufficient stems from the fact that our main results are independent of the precise value of $\Omega$ as far as $\Gamma_0 \ll \Omega \ll \omega_0$, which we verified explicitly by repeating calculations for several different $\Omega$. In addition, impurity oscillations influence light scattering via Doppler shifts $\propto \Delta\vec{k} \cdot \vec{v}$, where  $\Delta\vec{k}$ is the scattering vector and $\vec{v}$ is the instantaneous velocity of impurity. This quantity is random already for impurity oscillations with a fixed $\Omega$ because directions of $\Delta\vec{k}$ and $\vec{v}$ as well as the magnitude of $\vec{v}$ are random. Introducing a distribution of $\Omega$ would only modify the statistical properties of Doppler shifts without qualitatively modifying the physics of the problem. Finally, our main conclusion is that impurity oscillations do not impede observation of Anderson localization of light. This conclusion holds for any $\Omega$ in the range $\Gamma_0 \ll \Omega \ll \omega_0$ and thus it is unlikely to break down for a distribution of $\Omega$ in the same range.

\section*{Acknowledgements}

The work of IMS was supported by the Foundation for the Advancement of Theoretical Physics and Mathematics ``BASIS''.
Calculations were performed using the computing resources of the supercomputer center of Peter the Great St. Petersburg Polytechnic University.
(http://www.spbstu.ru).

\appendix

\section{Scattering mean free path}
\label{sec:mfp}

In the present work, we use the word ``transparent'' to characterize the medium in which oscillating impurity atoms are embedded. This word implies that all types of light scattering by the matrix are negligible. For a fiber-optics quality glass, for example, the strongest scattering process would be Rayleigh scattering, with a scattering mean free path exceeding ten kilometers \cite{jacobsen71}, which is many orders of magnitude larger than sample sizes that we have in mind in the present work. Other types of scattering and, in particular inelastic spontaneous Raman and Brillouin scattering due to molecular vibrations and thermal phonons, respectively, are typically orders of magnitude weaker \cite{brown87}, which makes them negligible as well. Stimulated scattering processes can be much more efficient but require strong enough intensity of light and are thus beyond the scope of the present work that is limited to linear, one-photon physics. Thus, oscillating impurity atoms constitute the only source of scattering in the considered system.

\begin{figure}[t]
\hspace*{3mm}
\includegraphics[width=\columnwidth]{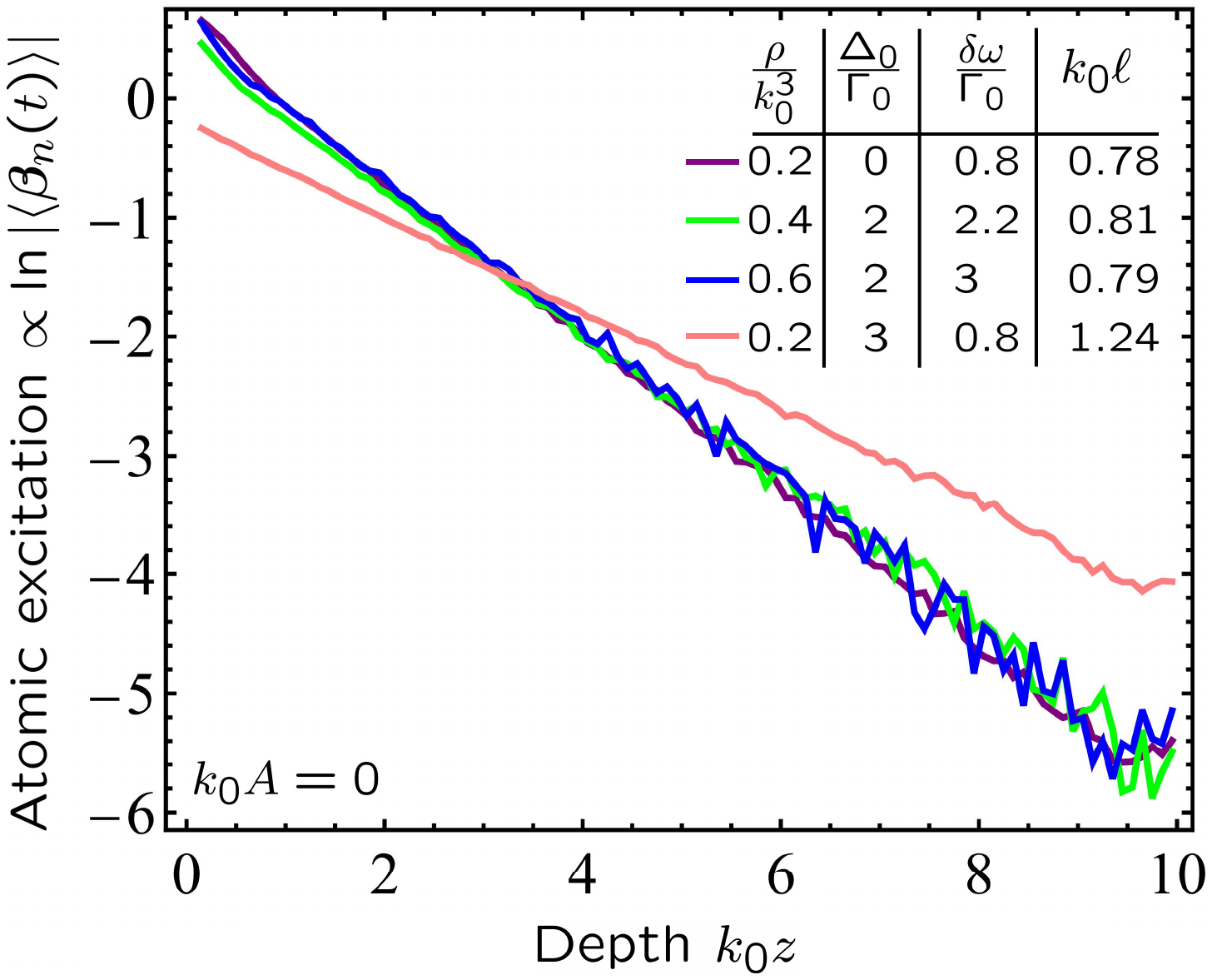}
\vspace*{-7mm}
\caption{\label{fig_mfp}
Ensemble average of the probability amplitude for the atom $n$ to be in the excited state as a function of depth $z$ inside the sample. Colors of curves correspond to those in Figs.\ 2 and 3. The table in the inset shows the scattering mean free paths $\ell$ obtained from negative exponential fits (\ref{mfp}) to the data.}
\end{figure}

To determine the scattering mean free path due to scattering of light by impurity atoms,
we use Eq.\ (\ref{iteration}) to compute the ensemble and time average of the probability amplitude $\bm{\beta}_n(t)$ for the atom at a position $\vec{r}_n$ to be in the excited state. Because results become independent from the amplitude of impurity atom oscillations for $k_0 A \ll 1$, we focus on the case of $k_0 A = 0$. Figure \ref{fig_mfp} shows $|\langle \bm{\beta}_n(t) \rangle|$ in the logarithmic scale for several sets of parameters. We observe a clear exponential decay of all curves with depth $z$. Because $\bm{\beta}_n(t) \propto \vec{E}(\vec{r}_n,t)$, this decay allows us to estimate the scattering mean free path $\ell$ of light by fitting the numerical data of Fig.\ \ref{fig_mfp} to the expected behavior \cite{sheng06}
\begin{equation}
\langle \bm{\beta}_n(t) \rangle \propto e^{-z/2\ell}
\label{mfp}
\end{equation}
The resulting values of $\ell$ are listed in the inset of Fig.\ \ref{fig_mfp}.

It should be kept in mind that Fig.\ \ref{fig_mfp} is obtained for circularly polarized incident light quasi-resonant with one of the transitions $|g \rangle \to |e_{\pm 1} \rangle$ that has a dipole moment $\vec{d}_{e_{\pm 1} g}$ parallel to the polarization vector of the incident wave.   Light of the same frequency but with opposite circular polarization would not be scattered by the impurity atoms at all, which would correspond to $\ell \to \infty$. We have already discussed such `transparency channels' in supplemental material of Ref.\ \cite{skip15prl}. Their existence compromises neither strong scattering of light with opposite helicity nor spatial localization of collective atomic states discussed in the main text.

\section{Inhomogeneous broadening in the independent scattering approximation}
\label{approx}

In the independent scattering approximation, the self-energy is \cite{sheng06,kaiser09}
\begin{eqnarray}
\begin{aligned}
\Sigma(\omega,\omega_m') &= \frac{4\pi\rho}{k_0} \times
\frac{1}{2(\omega-\omega_m') /\Gamma_0 + i}
\end{aligned}
\label{self}
\end{eqnarray}
We assume that the atomic resonance frequency $\omega_m'$  is subject to inhomogeneous broadening due to (i) strong local fields in the transparent host medium and (ii) strong dipole-dipole interactions between nearby atoms. The resulting distribution of $\omega_m'$ is approximated by a Gaussian:
\begin{eqnarray}
p(\omega_m') &=& \frac{1}{\sqrt{2\pi}\Delta} \exp\left[ - \frac{(\omega_m' - \omega_m)^2}{2 \Delta^2} \right]
\label{gauss}
\end{eqnarray}
where $\omega_m$ is the resonance frequency in the absence of broadening. The variance of the distribution (\ref{gauss}) is given by a sum of two contributions:
\begin{eqnarray}
\Delta^2 = \Delta_0^2 + \langle \Delta_{\text{dd}}^2 \rangle
\label{delta2}
\end{eqnarray}
First, local fields in the host crystal yield random frequency shifts that we denote by $\Delta_n$, with a variance $\Delta_0^2$. Second, the typical frequency shifts $\Delta_{\text{dd}}$ induced by dipole-dipole interactions are of the order of $1/(k_0 \Delta r)^3 \propto \rho/k_0^3$, where we used the fact that the typical distance between nearby atoms is $\Delta r = \rho^{-1/3}$. To obtain a quantitative estimate, we recall that in a strong magnetic field, the eigenvalues of the Hamiltonian (1), yielding eigenfrequencies $\omega$ near resonances $\omega_m $ ($m = \pm 1$), can be obtained by diagonalizing a $N \times N$ matrix \cite{skip15prl,skip18prl}
\begin{eqnarray}
\begin{aligned}
(\hat{H}_{\text{eff}})_{nn'} &= \left(i \mp 2\frac{\Delta_B}{\Gamma_0} \right) \delta_{nn'}
+ (1 - \delta_{nn'})  \frac{3}{2} \frac{e^{i k_0 \Delta r_{nn'}}}{k_0 \Delta r_{nn'}}
\\
&\times
\left[P(i k_0 \Delta r_{nn'}) + Q(i k_0 \Delta r_{nn'}) \frac{\sin^2 \theta_{nn'}}{2} \right]\;\;\;\;\;\;\;\;
\end{aligned}
\label{gm}
\end{eqnarray}
where $\theta_{nn'}$ is the angle between the vector $\Delta \vec{r}_{nn'} = \vec{r}_n - \vec{r}_{n'}$ and the quantization axis $z$. To be specific, let us consider $m=1$. For two atoms ($N = 2$) at a distance $\Delta\vec{r} = \vec{r}_2 - \vec{r}_1$ the two eigenvalues of the matrix (\ref{gm}) are
\begin{eqnarray}
\begin{aligned}
E_{\pm} &= \left(i - 2 \frac{\Delta_B}{\Gamma_0} \right) \pm \frac{3}{2} \frac{e^{i k_0 \Delta r}}{k_0 \Delta r}
\\
&\times
\left[P(i k_0 \Delta r) + Q(i k_0 \Delta r) \frac{\sin^2 \theta}{2} \right]
\end{aligned}
\label{gm2}
\end{eqnarray}
and frequency shifts with respect to $\omega_1 = \omega_0 + \Delta_B$ are $(\Delta_{\text{dd}})_{\pm} = -(\Gamma_0/2)\text{Re} E_{\pm} - \Delta_B$. Averaging over the two eigenvalues and over $\theta$ yields
\begin{eqnarray}
\langle \Delta_{\text{dd}} \rangle &=& 0
\label{dav}
\end{eqnarray}
\begin{eqnarray}
\begin{aligned}
&\langle \Delta_{\text{dd}}^2 \rangle =
\frac{3 \Gamma_0^2}{160} \left(\frac{\rho}{k_0^3} \right)^{\frac23}
\left\{3 \left(\frac{\rho}{k_0^3} \right)^{\frac43}+\left(\frac{\rho}{k_0^3} \right)^{\frac23}
\right. \\
&\left.+\left[3 \left(\frac{\rho}{k_0^3} \right)^{\frac43}-5 \left(\frac{\rho}{k_0^3} \right)^{\frac23}+7\right] \cos
   \left[ 2 \left(\frac{\rho}{k_0^3} \right)^{-\frac{1}{3}}  \right] \right.
\\
&+\left. \left[ 6 \left(\frac{\rho}{k_0^3} \right) -2 \left(\frac{\rho}{k_0^3} \right)^{\frac13} \right] \sin \left[ 2 \left(\frac{\rho}{k_0^3} \right)^{-\frac{1}{3}} \right]+7\right\}
\\
&\to \frac{9 \Gamma_0^2}{80} \left(\frac{\rho}{k_0^3} \right)^2
\end{aligned}
\label{dvar}
\end{eqnarray}
where in the last line we have taken the limit $\rho/k_0^3 \gg 1$.

The average self-energy is
\begin{eqnarray}
\langle \Sigma(\omega) \rangle &=&
\int\limits_{-\infty}^{\infty} d\omega_m' p(\omega_m') \Sigma(\omega,\omega_m')
\label{selfav}
\end{eqnarray}
The effective complex wave number is
\begin{eqnarray}
\sqrt{k_0^2 - \langle \Sigma(\omega) \rangle} &=& k_{\text{eff}} + \frac{i}{2\ell}
\label{keff}
\end{eqnarray}
and hence
\begin{eqnarray}
k_{\text{eff}} &=& \text{Re} \sqrt{k_0^2 - \langle \Sigma(\omega) \rangle}
\label{k1}
\\
\ell &=& \frac{1}{2\; \text{Im} \sqrt{k_0^2 - \langle \Sigma(\omega) \rangle}}
\label{l1}
\end{eqnarray}
Ioffe-Regel criterion of localization is
\begin{eqnarray}
k_{\text{eff}} \ell &=& \frac{\text{Re} \sqrt{k_0^2 - \langle \Sigma(\omega) \rangle}}{2\; \text{Im} \sqrt{k_0^2 - \langle \Sigma(\omega) \rangle}}
= (k_{\text{eff}} \ell)_c \sim 1
\label{ir}
\end{eqnarray}
Comparison of contour plots of this equation with the localization phase diagram following from numerical calculations is shown in Fig.\ \ref{fig_phase} for $(k_{\text{eff}} \ell)_c = 0.6$.

\newpage


%

\end{document}